\documentclass[%
aps, %
prb, 
twocolumn, 10pt, 
a4paper, %
eqsecnum, 
amsfonts, amsmath, amssymb,
floatfix]{revtex4-2}
\usepackage[sort&compress]{natbib}
\usepackage{newtxtext}
\usepackage[varg]{newtxmath}
\usepackage{ascmac}
\usepackage{mathrsfs}
\usepackage{bm}
\usepackage{color}
\usepackage{graphicx}
\usepackage{bm}
\usepackage{braket}
\usepackage{mathrsfs}
\usepackage[normalem]{ulem}
\usepackage{hhline}
\usepackage{hyperref}
\hypersetup{%
	unicode=false, %
	bookmarksnumbered=true, %
	bookmarkstype=toc, %
	breaklinks=true, %
	colorlinks=true, %
	pdfborder={0 0 1}, %
	linkcolor=blue, %
	citecolor=blue, %
	urlcolor=blue, %
	pdftoolbar=true, %
	pdfmenubar=true, %
	pdffitwindow=false, %
	pdfstartview={FitH} %
}
\newcommand{\kB}{k_{\mathrm{B}}}
\newcommand{\me}{m_{\mathrm{e}}}

\newcommand{\eF}{\epsilon_{\mathrm{F}}}
\newcommand{\G}{\mathrm{G}}
\newcommand{\STT}{\mathrm{stt}}
\newcommand{\br}{\bm{r}}
\newcommand{\bk}{\bm{k}}

\newcommand{\bq}{\bm{q}}
\newcommand{\zi}{\mathrm{i}}
\newcommand{\dd}[1]{\mathrm{d} #1\,}
\renewcommand{\Im}{\mathrm{Im}\,}
\newcommand{\R}{\mathrm{R}}
\newcommand{\A}{\mathrm{A}}
\newcommand{\X}{\mathrm{X}}
\newcommand{\Y}{\mathrm{Y}}

\newcommand{\FI}{\mathrm{F1}}
\newcommand{\FII}{\mathrm{F2}}
\newcommand{\NI}{\mathrm{N1}}
\newcommand{\NII}{\mathrm{N2}}
\newcommand{\NIII}{\mathrm{N3}}

\begin{document}
%
%
%
%
%
\title{Field theoretical approach to spin torques: Slonczewski torques}
\date{\today}
\author{Junji Fujimoto}
\affiliation{Department of Electrical Engineering, Electronics, and Applied Physics, Saitama University, Saitama, 338-8570, Japan}
\email[E-mail address: ]{fujimoto.junji@gmail.com}

\begin{abstract}
The quantum field theoretical approach with the Kubo formula has successfully captured spin torques, such as spin-transfer torques and spin-orbit torques, for continuum systems.
We examine the field theoretical approach to current-induced spin-transfer torques in a magnetic junction system.
We first give a brief overview of the field theoretical approach to spin torques.
Then, we consider a five-layers system consisting of three nonmagnetic metal layers separated by two ferromagnetic metal layers and apply an electric field perpendicular to the layers.
We demonstrate that the Slonczewski-type spin-transfer torque, or shortly the Slonczewski torque, on the magnetizations in ferromagnetic layers is obtained by evaluating nonequilibrium electron spin density, based on the linear response theory with the Green function method.
The obtained coefficient of the Slonczewski torque has a quantum oscillation at absolute zero temperature, which has not been mentioned before.
A field-like torque accompanied by the Slonczewski torque is also evaluated.

\end{abstract}
\maketitle

\section{\label{sec:intro}Introduction}
The quantum field theory is a powerful tool for investigating various physical phenomena from particle physics to condensed matter physics, and provides us intuitive physical pictures of the phenomena.
In spintronics, the field theoretical approach has succeeded in capturing the spin torques~\cite{tatara2004,tatara2008,manchon2008}, the spin-motive forces~\cite{shibata2011}, the spin pumping~\cite{ohnuma2014,tatara2017}, and more, whereas many spintronic phenomena remain yet to be studied based on quantum field theory.

Among others, spin torques in various continuum systems have been studied based on the quantum field theory.
Here, we refer to the spin torques as torques of the nonequilibrium conduction electron spin on the magnetization through the $sd$-type exchange interaction between the conduction electron spin and the magnetization.
Current-induced spin-transfer torques are one of the spin torques and arise by applying an electric current in ferromagnetic metals with spatially-varying magnetic textures, such as magnetic domain walls.
The current-induced spin torques have been extensively investigated based on the motivation for magnetization manipulation by electrical means.

Two typical current-induced spin-transfer torques in continuum ferromagnetic systems are known; one is called (adiabatic) spin-transfer torque given as $\bm{\tau}_{\STT} = (\bm{j}_{\mathrm{s}} \cdot \bm{\nabla}) \bm{m}$, and the other is called here the $\beta$ torque, which is given as $\bm{\tau}_{\beta} = \beta \bm{m} \times (\bm{j}_{\mathrm{s}} \cdot \bm{\nabla}) \bm{m}$, where $\bm{j}_{\mathrm{s}}$ is the spin-polarized current density, $\bm{m}$ is the local unit magnetization vector, and $\beta$ is a coefficient determined by spin relaxation processes and/or nonadiabaticity in the systems.
In continuum systems, the magnetization may have spatial and temporal dependences; $\bm{m} = \bm{m} (\br, t)$, which represents various magnetic textures.

On the other hand, the spin-transfer torque was firstly predicted in a magnetic junction system, based on quantum mechanics~\cite{slonczewski1996,berger1996,bazaliy1998} and is currently called Slonczewski-type spin-transfer torque, or shortly Slonczewski torque.
Slonczewski considered a five-layer system consisting of three nonmagnetic metals separated by two ferromagnetic metal layers whose magnetizations are noncollinear and apply an electric current perpendicular to the layers, which results in magnetization dynamics due to the spin transfer torque.
Here, we express the magnetizations in two magnetic layers as $\bm{M}_1$ and $\bm{M}_2$, and then the Slonczewski torque on $\bm{M}_i$ with $i = 1, 2$ is given as $\bm{T}_i = c \bm{M}_i \times (\bm{m}_1 \times \bm{m}_2)$, where $c$ is a coefficient determined by the applied electric current and its spin polarization~\cite{slonczewski1996}, and $\bm{m}_i = \bm{M}_i / |\bm{M}_i|$ ($i = 1, 2$).

The actual connection between the (adiabatic) spin-transfer torque $\bm{\tau}_{\STT}$ and the Slonczewski torque $\bm{T}_i$ has been unclear, whereas one can expect that both torques should be equivalent.
In the limiting case of the almost collinear magnetizations with the same lengths $|\bm{M}_i| = M$ ($i = 1, 2$), we can see that the Slonczewski torque is reduced to a $\bm{\tau}_{\STT}$-like torque, by setting $\bm{m}_2 = \bm{m}_1 + \delta \bm{m}$, where $\delta \bm{m}$ is small compared to $\bm{m}_1$ and perpendicular to $\bm{m}_1$, which yields $\bm{T}_1 = c \bm{M}_1 \times (\bm{m}_1 \times \bm{m}_2) = - c M \delta \bm{m}$.
Here, since applied electric current direction is perpendicular to the layers, say the $x$ direction, the (adiabatic) spin-transfer torque is described as $\bm{\tau}_{\STT} = j_{\mathrm{s}} \partial_x \bm{m}$ and discretized as $\partial_x \bm{m} = \{ \bm{m} (x + \delta x) - \bm{m} (x) \} / \delta x$, which results in $\bm{T}_1 \propto \bm{\tau}_{\STT}$ by setting $\bm{m} (x + \delta x) = \bm{m}_2$ and $\bm{m} (x) = \bm{m}_1$.
However, the above discussion is not valid in general noncollinear cases with different magnetization lengths, which are of our interest.
Further, the $\beta$ torque is known to be essential for magnetization dynamics in continuum systems, while the torque is sometimes disregarded in junction systems.
It may be worth evaluating the counterpart of the $\beta$ torque in the junction system.
Here, we call $\bm{T}_i$ the damping-like Slonczewski torque and the counter part of the $\beta$ torque in the junction systems, $\bm{T}'_1 = c' \bm{M}_1 \times \bm{m}_2$ and $\bm{T}'_2 = c' \bm{M}_2 \times \bm{m}_1$, the filed-like Slonczewski torque, where $c'$ is a coefficient different from $c$.

In this paper, we examine the field theoretical approach to Slonczewski torques.
First, following Slonczewski, we consider the five-layers system consisting of three nonmagnetic metal layers separated by two ferromagnetic metal layers and apply an electric field perpendicular to the layers.
We demonstrate that the Slonczewski torques are obtained in general noncollinear cases by evaluating nonequilibrium electron spin density, based on the linear response theory with the Green function method.
The coefficient of the damping-like Slonczewski torque, $c$, has a spatial quantum oscillation like the Friedel oscillation at absolute zero temperature, which has not been mentioned before.
Moreover, the coefficient $c$ takes a different magnitude depending on the magnetizations in the magnetic layers.
We also discuss the filed-like Slonczewski torque in the junction system.
The coefficient of the field-like torque has the same magnitude with the opposite sign in the magnetic layers.
We find that the field-like torque has the saturation value for large thicknesses of the magnetic layers.
Although recent interest seems to move from the spin-transfer torques to the spin-orbit torques, which arise only in systems with strong spin-orbit coupling~(SOC), we discuss current-induced spin-transfer torques in magnetic junction systems without SOC in this paper.

\section{\label{sec:overview}Overview}
We here give a brief overview of magnetization dynamics and the field theoretical approach to spin torques.

\subsection{\label{sec:dynamics}Magnetization dynamics and spin torque}
Consider two systems of localized (classical) spins consisting of the magnetization and of conduction electrons, both of which are coupled through the $sd$-type exchange interaction~\footnote{Other interactions between the localized spin and the conduction electron spin, such as the RKKY interaction and the Dzyaloshinskii-Moriya interaction, possibly can arise novel types of spin torques, but we focus on the spin torques originating from the $sd$-type exchange interaction.}.
The total system is described by the following equation;
\begin{align}
\mathcal{L}
	& = \mathcal{L}_s + \mathcal{L}_e - \mathcal{H}_{sd}
\label{eq:L}
	,\end{align}
where $\mathcal{L}_s$ and $\mathcal{L}_e$ are the Lagrangians of the localized spin system and the conduction electron system, and $\mathcal{H}_{sd}$ is the $sd$-type exchange interaction between the localized spin and the conduction electron spin.
Here, we express the $i$-th localized spin degree of freedom as $\bm{S}_i$, whose length is fixed as $|\bm{S}_i| = S_i$, where $S_i$ is related to the saturated magnetization $M_{S,i}$ as $M_{S,i} = \gamma_e \hbar S_i / a^3$ with $\gamma_e$ being the gyromagnetic ratio and $a$ the length scale of coarse graining.

The dynamics of the localized spin $\bm{S}_i$ obeys classical mechanics and determined by the Euler-Lagrange equation with the constrain $\mathcal{C}_i = |\bm{S}_i|^2 - S_i^2 = 0$ and the phenomenological relaxation function $\mathcal{W}_s$;
\begin{align}
\frac{d}{d t} \left( \frac{\delta \mathcal{L}}{\delta \dot{\bm{S}}_i} \right)
	- \frac{\delta \mathcal{L}}{\delta \bm{S}_i}
	+ \lambda \frac{\partial \mathcal{C}_i}{\delta \bm{S}_i}
	& = \frac{\delta \mathcal{W}_s}{\delta \dot{\bm{S}}_i}
\label{eq:EL}
	,\end{align}
where $\lambda$ is the Lagrange multiplier.
Here, thermal average on the conduction electron system is implied.
Substituting Eq.~(\ref{eq:L}) into Eq.~(\ref{eq:EL}), we have
\begin{align}
	\frac{d}{d t} \left( \frac{\delta \mathcal{L}_s}{\delta \dot{\bm{S}}_i} \right)
	- \frac{\delta \mathcal{L}_s}{\delta \bm{S}_i}
	+ \frac{\delta \mathcal{H}_{sd}}{\delta \bm{S}_i}
	+ \lambda \bm{S}_i
	& = \frac{\delta \mathcal{W}_s}{\delta \dot{\bm{S}}_i}
,\end{align}
and then taking $\bm{S}_i \times$ to remove the Lagrange multiplier, we obtain
\begin{align}
	\bm{S}_i \times \frac{d}{d t} \left( \frac{\delta \mathcal{L}_s}{\delta \dot{\bm{S}}_i} \right)
	- \bm{S}_i \times \frac{\delta \mathcal{L}_s}{\delta \bm{S}_i}
	+ \hbar \bm{\tau}_i
	& = \bm{S}_i \times \frac{\delta \mathcal{W}_s}{\delta \bm{S}_i}
\label{eq:EOM}
,\end{align}
where $\bm{\tau}_i$ is the spin torque given by
\begin{align}
\bm{\tau}_i
	& = \bm{S}_i \times \left\langle \frac{1}{\hbar} \frac{\delta \mathcal{H}_{sd}}{\delta \bm{S}_i} \right\rangle
\label{eq:tau_e}
.\end{align}
Here, $\langle \, \cdots \rangle$ means the thermal average on the conduction electorn system.
Note that $\mathcal{L}_e$ does not contain $\bm{S}_i$, hence $\delta \mathcal{L}_e / \delta \bm{S}_i = 0$, and effects of the conduction electron on the localized spins are only through the $sd$-type exchange interaction.
We also note that the spin torque $\bm{\tau}_i$ does not cause the magnetization dynamics in global equilibrium (by definition of the equilibrium), which is understood as that the conduction electron spins align along the localized spins.
The spin torque $\bm{\tau}_i$ can induce the magnetization dynamics in nonequilibrium, such as (i) by applying an external electric field, (ii) by a temperature gradient, and (iii) by induced magnetization dynamics.
The spin torques of the case (i) is called the curren-induced spin torques~\cite{slonczewski1996,berger1996,tatara2004,zhang2004,duine2007,ralph2008,tatara2008,brataas2012,manchon2008,manchon2019}, the case (ii) is named the thermal spin torques~\cite{hatami2007,kohno2016}, and the case (iii) is related to the Gilbert damping torque due to the conduction electrons~\cite{kohno2006,tserkovnyak2006,duine2007,kohno2007}.

For the system of ferromagnetically-interacting localized spins, the terms in Eq.~(\ref{eq:EOM}) read
\begin{align}
\bm{S}_i \times \frac{d}{d t} \left( \frac{\delta \mathcal{L}_s}{\delta \dot{\bm{S}}_i} \right)
	= - \hbar \dot{\bm{S}}_i
, & \quad
\bm{S}_i \times \frac{\delta \mathcal{L}_s}{\delta \bm{S}_i}
	= - \hbar \bm{S}_i \times \gamma_e \bm{B}_i
,\end{align}
where $\bm{B}_i$ is the effective magnetic field (measured by the unit of Tesla) originating from the localized spin Hamiltonian $\mathcal{H}_s$, given as
\begin{align}
\gamma_e \bm{B}_i
	& = \frac{1}{\hbar} \frac{\partial \mathcal{H}_s}{\partial \bm{S}_i}
.\end{align}
The relaxation function in ferromagnets is given as $\mathcal{W}_s = (\hbar \alpha_{\G}/2 S) \sum_i \dot{\bm{S}}_i^2$, which leads to the phenomenological Gilbert damping torque
\begin{align}
\bm{S}_i \times \frac{\delta \mathcal{W}_s}{\delta \bm{S}_i}
	& = \frac{\hbar \alpha_{\G}}{S} \bm{S}_i \times \dot{\bm{S}}_i
,\end{align}
and then Eq.~(\ref{eq:EOM}) is found equivalent to Landau-Lifshitz-Gilbert equation with the spin torque,
\begin{align}
\dot{\bm{S}}_i
	& = \gamma_e \bm{S}_i \times \bm{B}_{i}
		- \frac{\alpha_{\G}}{S} \bm{S}_i \times \dot{\bm{S}}_i
		+ \bm{\tau}_i
.\end{align}
By solving this equation of motion, we see the magnetization dynamics.

We here emphasize that the spin torque is given as the effects of the conduction electrons on the magnetization dynamics, as mentioned above, and the above discussion is valid not only for continuum systems but for junction systems, the latter of which is of our interest.
For continuum systems, we usually take the continuum limit $\bm{S}_i \to \bm{S} (\br)$.
Note also that we can discuss the order parameters dynamics for the antiferromagnetic and ferrimagnetic systems, but they are out of scope in this paper.

\subsection{\label{sec:spin-torque}Spin torque in continuum and junction systems}
Next, we discuss the expressions of the spin torque in continuum and junction systems.
The $sd$-type exchange interaction $\mathcal{H}_{sd}$ is given by
\begin{align}
\mathcal{H}_{sd}
	& = - \sum_{i} \int \dd{\br} J_i (\br) \bm{S}_i \cdot \bm{s} (\br)
\label{eq:H_sd}
\end{align}
in general, where $J_i (\br)$ is the interaction strength between the $i$-th localized spin and electron spin, $\bm{s} (\br)$ is the conduction electron spin density (devided by $\hbar/2$).
The definition with Eq.~(\ref{eq:tau_e}) immediately leads to the following expression of the spin torque,
\begin{align}
\bm{\tau}_i
	& = - \frac{1}{\hbar} \bm{S}_i \times \left\langle \int \dd{\br} J_i (\br) \bm{s} (\br) \right\rangle
\label{eq:tau_sd}
.\end{align}
This equation indicates that the spin torque is an effective magnetic field due to the conduction electron spins coupling to the localized spin.

In the local interaction case; $J_i (\br) = J a^3 \delta (\br - \bm{R}_i)$, where $\bm{R}_i$ is the position of the $i$-th localized spin, and $\delta (\br)$ is the Dirac $\delta$-function, the spin torque is simply given as
\begin{align}
\bm{\tau}_i
	& = - \frac{J a^3}{\hbar} \bm{S}_i \times \langle \bm{s} (\bm{R}_i) \rangle
.\end{align}
In the continuum limit, the spin torque is shown as
\begin{align}
\bm{\tau}_i
	& \to \bm{\tau} (\br)
	= \frac{J_{sd}}{\hbar} \bm{m} (\br) \times \langle \bm{s} (\br) \rangle
,\end{align}
where $J_{sd} = S J a^3$ is the $sd$ exchange interaction constant, $\bm{m} (\br)$ is the unit magnetization which is antiparallel to the localized spin $\bm{S} (\br)$.

In this paper, we consider magnetic junction systems, where the magnetization of the $i$-th magnetic layer is assumed uniform in the layer.
In the manner of the definition~(\ref{eq:H_sd}), the $sd$-type exchange interaction is treated by the following approximation;
\begin{align}
J_i (\br)
	& = J \Theta_i (\br)
	= \left\{ \begin{array}{l l}
		J
	&	(\br \in \Omega_i)
	\\	0
	&	\text{(otherwise)}
	,\end{array}
	\right.
\end{align}
where $\Omega_i$ is the space of the $i$-th magnetic layer, and $\Theta_i (\br)$ is the step function defined by the above equation.
The spin torque in this case is obtained as
\begin{align}
\bm{\tau}_i
	& = - \frac{J}{\hbar} \bm{S}_i \times \left\langle \int_{\Omega_i} \dd{\br} \bm{s} (\br) \right\rangle
.\end{align}
We use this expression for calculating the Slonczewski torques.

\subsection{Linear response theory for spin torque}
Then, we see how to evaluate the spin torque in various nonequilibrium cases, based on the linear response theory.
In the previous subsection, we see the spin torque is defined by the effective magnetic field due to the conduction electron spin density~[Eq.~(\ref{eq:tau_sd})].
Hence, we can obtain the spin torque by evaluating the conduction spin density in nonequilibirum, and the evaluation is performed based on the linear response theory.

Below are spin torques known in continuum systems.

\subsubsection{\label{sec:CIST}Current-induced spin torques}
For the current-induced spin torque, the spin density is evaluated as
\begin{align}
\langle s^{\alpha} (\br) \rangle
	& = \int \dd{\br'} \chi_i^{\alpha} (\br, \br') E_i (\br')
\label{eq:response_E}
,\end{align}
where $\alpha$ and $i$ are the spin index and the direction index of the applied electric field, respectively.
From the linear response theory, the response coefficient $\chi_i^{\alpha} (\br, \br')$ is found to be obtained as
\begin{align}
\chi_i^{\alpha} (\br, \br')
	& = \lim_{\omega \to 0} \frac{K_i^{\alpha} (\br, \br'; \omega) - K_i^{\alpha} (\br, \br'; 0)}{\zi \omega}
\end{align}
with the spin-current correlation function $K_j^{\alpha} (\br, \br'; \omega)$,
\begin{align*}
K_i^{\alpha} (\br, \br'; \omega)
	& = \frac{\zi}{\hbar} \int_0^{\infty} \dd{t} e^{\zi (\omega + \zi 0) t}
		\left\langle [ s^{\alpha} (\br, t), j_i (\br') ] \right\rangle
,\end{align*}
where $[A, B] = A B - B A$ is the communicator, $s^{\alpha} (\br, t)$ is the spin density operator of the Heisenberg picture of the spin density, $j_i (\br)$ is the electric current density operator.
The correlation function $K_j^{\alpha} (\br, \br'; \omega)$ can be evaluated by using some techniques, such as the thermal Green function with the analytic continuation.
Note that it is possible to evaluate the nonequilibrium spin density $\langle s^{\alpha} (\br) \rangle$ based on the Keldysh Green function, by expanding the external force.

For a simple ferromagnetic metal with magnetization texture (without any SOCs), the nonequilibrium spin density is calculated as
\begin{align*}
\langle \bm{s} (\br) \rangle
	& = \frac{\hbar}{J_{sd}} \left\{
		\bm{m} (\br) \times (\bm{j}_{\mathrm{s}} \cdot \bm{\nabla}) \bm{m} (\br)
		+ \beta (\bm{j}_{\mathrm{s}} \cdot \bm{\nabla}) \bm{m} (\br)
	\right\}
,\end{align*}
which leads to the adiabatic spin-transfer torque and the $\beta$ torque, $\bm{\tau} = \bm{\tau}_{\STT} + \bm{\tau}_{\beta}$.
Here, $\bm{j}_{\mathrm{s}} = \sigma_{\mathrm{s}} \bm{E}$ is the spin-polarized current with the conductivity $\sigma_{\mathrm{s}} = \sigma_{\uparrow} - \sigma_{\downarrow}$, where $\sigma_{\uparrow}$ and $\sigma_{\downarrow}$ are the spin-resolved conductivities.
Note that, in the alternating current region, another type of spin torque arises~\cite{fujimoto2019}.

For the two-dimensional~(2D) system with the Rashba SOC, the electric current induces the spin polarization, which is known as the Edelstein effect and obtained from Eq.~(\ref{eq:response_E});
\begin{align}
\langle \bm{s} (\br) \rangle
	& \sim \lambda_{\R} \hat{z} \times \bm{j}_e
\label{eq:Edelstein}
,\end{align}
where $\hat{z}$ is assumed to be the direction of broken inversion symmetry and perpendicular to the 2D plane, $\lambda_{\R}$ is the Rashba SOC strength, and $\bm{j}_e$ is the uniform electric current.
This current-induced spin polarization leads to the Rashba spin-orbit torque~\cite{manchon2008,manchon2019}
\begin{align}
\bm{\tau}_{\R}
	& \sim \frac{\lambda_{\R} J_{sd}}{\hbar} \bm{m} \times ( \hat{z} \times \bm{j}_e )
\label{eq:RashbaSOT}
.\end{align}
The Rashba spin-orbit torque exists even when the magnetization is uniform and proportional to the SOC strength, hence different from the above spin-transfer torques.
Equation~(\ref{eq:Edelstein}) does not depend on the magnetization, but, since we consider the electron system coupling to the magnetization, another type of spin polarization may exist, such as
\begin{align}
\langle \bm{s} (\br) \rangle
	& = C \bm{m} \times ( \hat{z} \times \bm{j}_e )
\label{eq:field-likeSOT}
,\end{align}
which is actually found in the lattice model of 2D Rashba system, while $C = 0$ in the continuum model of the 2D Rashba system~\cite{fujimoto2015,ado2017}.

We should note that other types of current-induced spin torques are known in the presence both of magnetization textures and of SOC, which can be found in magnetic skyrmion systems~\cite{kurebayashi2019}.

\subsubsection{Thermal spin torques}
We see the thermal spin torques very briefly, which are spin torques induced by temperature gradients.
The response of the spin density to the temperature gradient is formally given as
\begin{align}
\langle s^{\alpha} (\br) \rangle
	& = \int \dd{\br'} \chi_i^{\alpha} (\br - \br') \bm{\nabla}'_i T (\br')
,\end{align}
where $\bm{\nabla}'_i$ indicates the gradient of the $i$ direction for the position $\br'$, and we assume the translational symmetry in the system we consider for simplicity.
Phenomenologically, discussions similar to the current-induced spin torques can be done, which leads to the thermal spin-transfer torques~\cite{hatami2007,kohno2016}.
To discuss more rigorously based on the Kubo formula, we have to introduce the fictional gravitational potentials that couple to the heat density or heat current density~\cite{luttinger1964,tatara2015}, since the temperature gradient is not a mechanical force but a statistical force, and the Kubo formula is only valid for the mechanical force.
The resultant expressions of the spin torques are same as phenomenologically derived forms~\cite{hatami2007,kohno2016} for thermal spin-transfer torques.
Thermal spin-orbit toruqes are also discussed in Refs.~\cite{vanderbijl2014,fujimoto2015}.

\subsubsection{Gilbert damping torque due to conduction electrons}
At the end of this overview, we would like to mention the Gilbert damping torque due to the conduction electrons.
The physical picture of this torque is as follows: the nonequilibrium spin density is induced by the magnetization dynamics, and conversely, the spin density acts as a spin torque on the magnetization.

The nonequilibrium spin density giving rise to the Gilbert damping torque is obtained from the linear response;
\begin{align}
\langle s^{\alpha} (\br) \rangle
	& = \int \dd{\br'} \chi^{\alpha \beta} (\br, \br') \dot{m}^{\alpha} (\br')
,\end{align}
where the response coefficient $\chi^{\alpha \beta} (\br, \br')$ is given as
\begin{align}
\chi^{\alpha \beta} (\br, \br')
	& = \frac{Q^{\alpha \beta} (\br, \br'; \omega) - Q^{\alpha \beta} (\br, \br'; 0)}{\zi \omega}
\end{align}
with the spin-spin correlation function
\begin{align*}
Q^{\alpha \beta} (\br, \br'; \omega)
	& = \frac{\zi J_{sd}}{\hbar} \int_0^{\infty} \dd{t} e^{\zi (\omega + \zi 0) t}
		\left\langle [ s^{\alpha} (\br, t), s^{\beta} (\br') ] \right\rangle
.\end{align*}
In the presence of the spin relaxation of the conduction electrons, the response coefficient is nonzero and
\begin{align}
\chi^{\alpha \beta}
	& = \frac{\hbar}{J_{sd}} \tilde{\alpha}_{\G} \delta^{\alpha \beta}
,\end{align}
where $\tilde{\alpha}_{\G}$ is a coefficient determined by the spin relaxation mechanism.
This nonequilibrium spin density arises the Gilbert damping torque as
\begin{align}
\bm{\tau}_{\G}
	& = \tilde{\alpha}_{\G} \bm{m} \times \dot{\bm{m}}
.\end{align}
Phenomenologically, the Gilbert damping torques originating from the strong SOC is captured by the Fermi surface breathing effect~\cite{kambersky1970,hillebrands2003}.

\section{\label{sec:model}five-layers system}
\subsection{Model}
\begin{figure}[htpb]
\centering
\includegraphics[width=\linewidth]{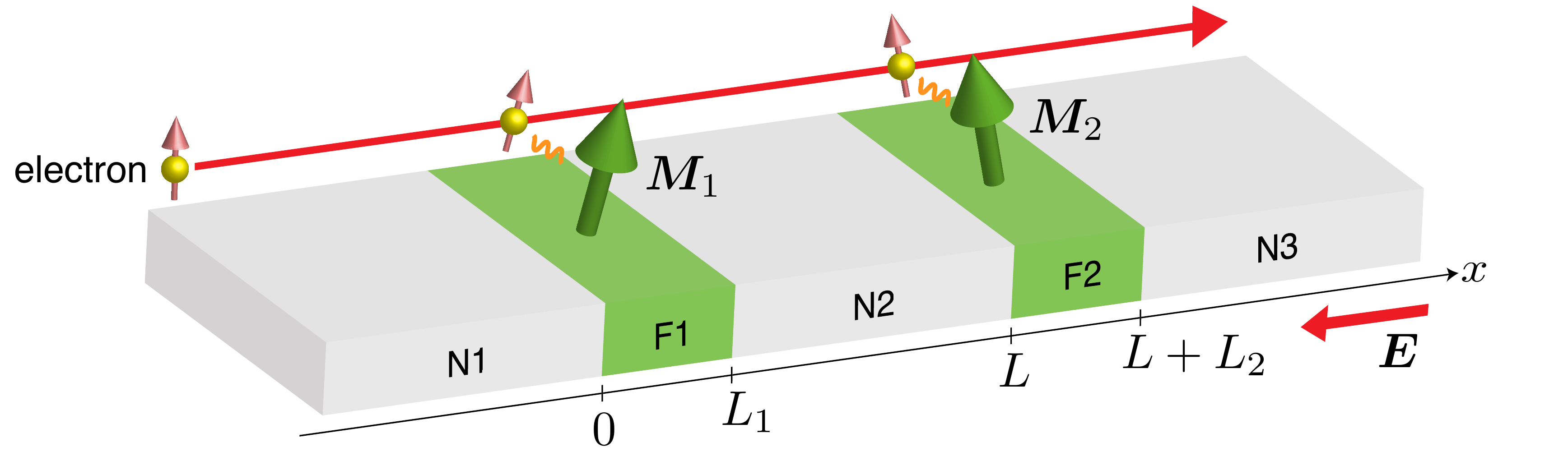}
\caption{\label{fig:5layer}Schematics of the five-layers system.}
\end{figure}

Then, we see the correspondence of Slonczewski torque to the (adiabatic) spin-transfer torque.
The framework of this calculation is based on the previous section.
We first present the model which we consider in this paper.
Following Slonczewski, we consider the five-layers system which consists of two ferromagnetic metal layers (denoted by $\FI$ and $\FII$) sandwiched between three nonmagnetic metals~(denoted by $\NI$, $\NII$, and $\NIII$).
The thicknesses of $\FI$ and $\FII$ are denoted by $L_1$ and $L_2$, and the magnetizations are $\bm{M}_1$ and $\bm{M}_2$, respectively.
The cross section of the layers is given by $A$, and the distance between $\FI$ and $\FII$ is shown by $\delta = L - L_1$.
Figure~\ref{fig:5layer} depicts the system that we consider here, and the $x$ direction is taken perpendicular to the layers.

The total Lagrangian density is given as
\begin{align}
\mathcal{L}
	& = \mathcal{L}_{\bm{M}_1} + \mathcal{L}_{\bm{M}_2} - \mathcal{H}_e - \mathcal{H}_{\mathrm{ext}}
\label{eq:L_5layer}
,\end{align}
where $\mathcal{L}_{\bm{M}_1}$ and $\mathcal{L}_{\bm{M}_2}$ are the Lagrangians of the magnetizations, which are not specified in this paper since we are focusing on the spin torques. The third term of Eq.~(\ref{eq:L_5layer}) is the Hamiltonian of the conduction electron in the five-layers system given as
\begin{align}
\mathcal{H}_e
	& = - \frac{\hbar^2\bm{\nabla}^2}{2 \me} - J_1 \tilde{\bm{m}}_1 (\br) \cdot \bm{\sigma}  - J_2 \tilde{\bm{m}}_2 (\br) \cdot \bm{\sigma}
\label{eq:H_e}
,\end{align}
which consists of the kinetic energy and the $sd$-type exchange interaction with the magnetizations.
Here, $\me$ is the electron mass, $J_i$ is the exchange interaction strength, $\bm{\sigma} = (\sigma^x, \sigma^y, \sigma^z)$ is the Pauli matrix, and
\begin{subequations}
\begin{align}
\tilde{\bm{m}}_1 (\br)
	& = \bm{m}_1 \Theta (x) \Theta (L_1 - x)
, \\
\tilde{\bm{m}}_2 (\br)
	& = \bm{m}_2 \Theta (x - L) \Theta (L + L_2 - x)
\end{align}%
\end{subequations}
with the Heaviside step function $\Theta (x)$ and the unit magnetization vectors $\bm{m}_i = \bm{M}_i / |\bm{M}_i|$ ($i = 1, 2$).
The meaning of $sd$-type exchange interction in Eq.~(\ref{eq:H_e}) is slightly different from that in Eq.~(\ref{eq:H_sd}); the exchange interaction strength does not depend on its position, while Eq.~(\ref{eq:H_sd}) is a position-dependent interaction strength.
However, the difference is not important, since we can rewrite it as $J_1 \tilde{\bm{m}}_1 (\br) = J_1 (\br) \tilde{\bm{m}}_1$ with $J_1 (\br) = J \Theta (x) \Theta (L_1 - x)$, and so on.
Note also that we treat the conduction electron is described by the single Hamiltonian~(\ref{eq:H_e}), which may be too much simplification to perform the quantitative evaluation of Slonczewski torques, but the simplification is valid for the qualitative discussion as seen below.
The last term of the Lagrangian~(\ref{eq:L_5layer}) indicates that the external electric field induces the electric current, which is given by
\begin{align}
\mathcal{H}_{\mathrm{ext}}
	& = - \bm{J}_e \cdot \bm{A} (t)
,\end{align}
where $\bm{J}_e$ is the electric current, and $\bm{A} = A_0 \hat{x} e^{- \zi \omega t}$ is the vector potential which induces the electric field 
\begin{align}
\bm{E}
	& = \zi \omega A_0 \hat{x} e^{- \zi \omega t}
\label{eq:E}
\end{align}
with $\omega \to 0$.

In the second quantization representation, the model is given as
\begin{align}
\mathcal{H}_e
	& = \sum_{\bk} \frac{\hbar^2 k^2}{2 \me} c^{\dagger}_{\bk} c^{}_{\bk}
		- V \sum_{\bq} \bm{s} (- \bq) \cdot (J_1 \tilde{\bm{m}}_1 (\bq) + J_2 \tilde{\bm{m}}_2 (\bq))
,\end{align}
where $c_{\bk}^{(\dagger)}$ is the annihilation (creation) operator of electron with the wavevector $\bk$,
\begin{align}
\tilde{\bm{m}}_i (\bq)
	& = \frac{1}{V} \int \dd{\br} \tilde{\bm{m}}_i (\br) e^{- \zi \bq \cdot \br}
,\end{align}
and the spin density $\bm{s} (\bq)$ and the charge current $\bm{J}_e$ are given by
\begin{align}
\bm{s} (\bq)
	= \frac{1}{V} \sum_{\bk} c^{\dagger}_{\bk - \bq} \bm{\sigma} c^{}_{\bk}
, \qquad
\bm{J}_e
	= - e \sum_{\bk} \frac{\hbar \bk}{\me} c^{\dagger}_{\bk} c^{}_{\bk}
,\end{align}
where $V = A W$ is the volume of the system with system length $W$, and $e \,(> 0)$ is the elementary charge.

\subsection{\label{sec:perturbation}Linear response theory}
For the above-mentioned system, we evaluate the Slonczewski torques.
The fundamental way of the evaluation is shown in Sec.~\ref{sec:overview} (\ref{sec:dynamics}, \ref{sec:spin-torque}, and \ref{sec:CIST}).
Firstly, the magnetization dynamics is described by the Landau-Lifshitz-Gilbert equation,
\begin{align}
	\frac{\mathrm{d} \bm{M}_i}{\mathrm{d} t}
		& = - \gamma \bm{M}_i \times \bm{H}_i + \frac{\alpha_{\mathrm{G}}}{M_S} \bm{M}_i \times \frac{\mathrm{d} \bm{M}_i}{\mathrm{d} t}
		+ \bm{T}_i
	\label{eq:LLG}
,\end{align}
with $i = 1, 2$, where the first term of the right hand side arises the precession motion, the second term gives rise to the damping motion to the direction of the equilibrium state, and the last term indicates the spin torque.
Note that $\bm{H}_i$ is determined by the Lagrangian of the magnetization $\bm{M}_i$, and $\alpha_{\mathrm{G}}$ is the Gilbert damping constant.
The spin torque is given by
\begin{align}
\bm{T}_i
	& = - \frac{J_i}{\hbar} \bm{M}_i \times \left\langle \int_{\Omega_i} \dd{\br} \bm{s} (\br) \right\rangle_{\mathcal{H}_e + \mathcal{H}_{\mathrm{ext}}}
\label{eq:spin-torque}
,\end{align}
where $\Omega_i$ is the space of the magnetic layer ($i = 1, 2$), and $\bm{s} (\br)$ is the spin density operator defined by $\bm{s} (\br) = \psi^{\dagger} (\br) \bm{\sigma} \psi (\br)$ with the electron field operator $\psi^{(\dagger)} (\br)$, which can be expanded by the annihilation (creation) operator $c_{\bk}^{(\dagger)}$ as
\begin{align}
\psi (\br)
	= \frac{1}{\sqrt{V}} \sum_{\bk} c_{\bk} e^{\zi \bk \cdot \br}
, & \quad
\psi^{\dagger} (\br)
	= \frac{1}{\sqrt{V}} \sum_{\bk} c_{\bk}^{\dagger} e^{- \zi \bk \cdot \br}
.\end{align}

To evaluate the spin torque, we now consider that the response of the spin to the external field, which is expressed as
\begin{align}
\langle s^{\alpha} (\br) \rangle_{\mathcal{H}_e + \mathcal{H}_{\mathrm{ext}}}
	& = \chi^{\alpha}_{j} (\br) E_j
\label{eq:response_s}
.\end{align}
Here, we first consider a general case of applied electric field and calculate the response coefficient $\chi_j^{\alpha} (\br)$.
Then, we will assume the specific configuration given in Eq.~(\ref{eq:E}) and obtain the damping-like Slonczewski torque.
The response coefficient $\chi^{\alpha}_{j} (\br)$ is evaluated by using the linear response theory;
\begin{align}
\chi_{j}^{\alpha} (\br)
	& = \lim_{\omega \to 0} \frac{ K^{\alpha}_j (\omega; \br) - K_j^{\alpha} (0; \br) }{\zi \omega}
,\end{align}
where $K^{\alpha}_j (\omega)$ is obtained from the corresponding Matsubara correlation function given by
\begin{align}
\mathcal{K}^{\alpha}_j (\zi \omega_{\lambda}; \br)
	& = \int_0^{\beta} \mathrm{d}\tau e^{\zi \omega_{\lambda} \tau} \langle \mathrm{T}_{\tau} \{ s^{\alpha} (\br, \tau) J_{e, j} \} \rangle_{\mathcal{H}_e}
\end{align}
with the analytic continuation $\zi \omega_{\lambda} \to \hbar \omega + \zi 0$;
\begin{align}
K_j^{\alpha} (\omega; \br)
	& = \mathcal{K}_j^{\alpha} (\hbar \omega + \zi 0; \br)
.\end{align}
Note that $\beta = 1 / \kB T$ is the inverse temperature, $\omega_{\lambda} = 2 \pi \lambda \kB T$ is the bosonic Matsubara frequency with $\lambda$ being integer, $\tau$ is the imaginary time (in the energy unit), $\mathrm{T}_{\tau}$ is the imaginary-time ordering operator, $s^{\alpha} (\br, \tau)$ is the Heisenberg operator in imaginary time, and $\langle \cdots \rangle_{\mathcal{H}_e}$ is the thermal average on the Hamiltonian $\mathcal{H}_e$.

In the Fourier space, Eq.~(\ref{eq:response_s}) is written as
\begin{align}
\langle s^{\alpha} (\bq) \rangle_{\mathcal{H}_e + \mathcal{H}_{\mathrm{ext}}}
	& = \chi^{\alpha}_{j} (\bq) E_j
\label{eq:response_s_in_k}
,\end{align}
where
\begin{align}
\chi^{\alpha}_{j} (\bq)
	& = \frac{1}{V} \int \dd{\br} \chi^{\alpha}_{j} (\br) e^{- \zi \bq \cdot \br}
\label{eq:chi_q}
.\end{align}
The Matsubara correlation function in the Fourier space is obtained as
\begin{align}
\mathcal{K}^{\alpha}_j (\zi \omega_{\lambda}; \bq)
	& = \int_0^{\beta} \mathrm{d}\tau e^{\zi \omega_{\lambda} \tau} \langle \mathrm{T}_{\tau} \{ s^{\alpha} (\bq, \tau) J_{e, j} \} \rangle_{\mathcal{H}_e}
.\end{align}

\subsection{Damping-like Slonczewski torque}
\begin{figure}[thpb]
\centering
\includegraphics[width=\linewidth]{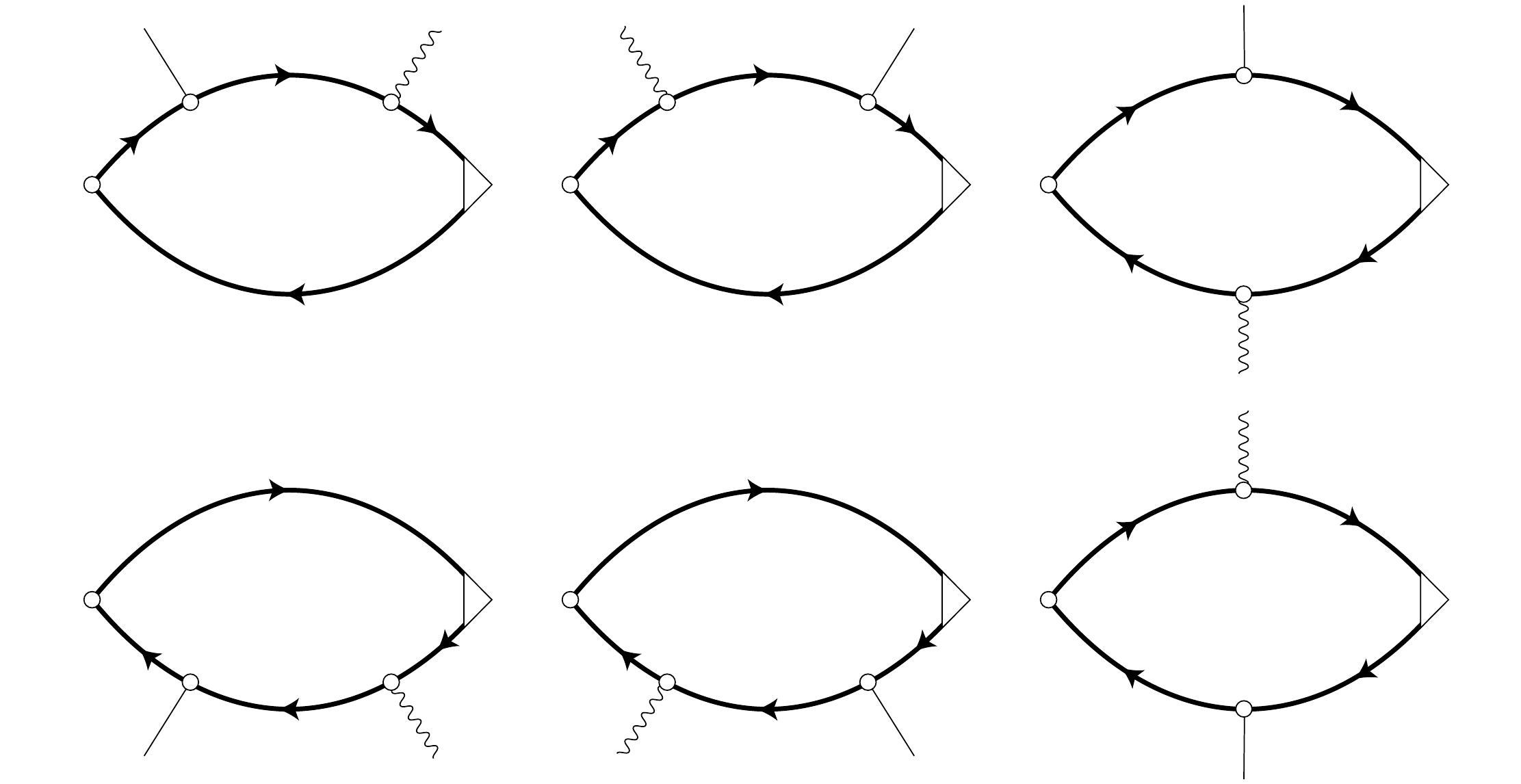}
\caption{\label{fig:diagram_damping-like}Feynman diagrams for the damping-like Slonczewski torque.
The solid lines with arrows represent the electron Green functions, the solid and wavy lines without arrows are $sd$-type exchange interactions with $\tilde{\bm{m}}_1 (\bq)$ and $\tilde{\bm{m}}_2 (\bq)$, respectively, the circle and triangle symbols indicate the spin and velocity vertexes, respectively.}
\end{figure}
To evaluate the damping-like Slonczewski torque $\bm{T}_i$, we consider the uniform component ($\bq = 0$) in Eq.~(\ref{eq:response_s_in_k}), because the $\bq \neq 0$ components oscillate spatially and are mainly canceled by integrating in the magnetic layer.
Then, we expand the response coefficient by the $sd$-type exchange interactions up to the first orders of $\bm{m}_1$ and $\bm{m}_2$ (see Fig.~\ref{fig:diagram_damping-like} for the corresponding Feynman diagrams), which results in
\begin{align}
\mathcal{K}^{\alpha}_j (\zi \omega_{\lambda})
	& = \mathcal{K}^{\alpha}_j (\zi \omega_{\lambda}; \bq = 0)
\notag \\ & \hspace{-3em}
	= - \frac{2 \zi e J_1 J_2}{\beta} \sum_{n, \bq'} \varphi_j (\zi \epsilon_n^{+}, \zi \epsilon_n; - \bq')
		\left\{ \tilde{\bm{m}}_1 (\bq') \times \tilde{\bm{m}}_2 (-\bq') \right\}^{\alpha}
\label{eq:K_1}
,\end{align}
where $\epsilon_n = (2n + 1) \pi \kB T$ and $\zi \epsilon_n^{+} = \zi \epsilon_n + \zi \omega_{\lambda}$ are the fermionic Matsubara frequency with $n$ being integer, and
\begin{align}
\varphi_j (\zi \epsilon_n^{+}, \zi \epsilon_n; - \bq)
	& = \frac{2}{V} \sum_{\bk} \biggl(
		\frac{\hbar k_j}{\me} g^{+}_{\bk+\bq} (g^{+}_{\bk})^2 g^{}_{\bk}
\notag \\ & \hspace{3em}
		+ \frac{\hbar k_j}{\me} g^{+}_{\bk} g^{}_{\bk+\bq} (g^{}_{\bk})^2
		\notag \\ & \hspace{3em}
		+ \frac{\hbar (k_j + q_j)}{\me} g^{+}_{\bk} g^{}_{\bk} g^{+}_{\bk+\bq} g^{}_{\bk+\bq}
	\biggr)
.\end{align}
Here, $g^{+}_{\bk} = g^{}_{\bk} (\zi \epsilon_n^{+})$, $g^{}_{\bk} = g^{}_{\bk} (\zi \epsilon)$ are the Matsubara Green functions of electrons.
By the inverse Fourier transformation, Eq.~(\ref{eq:K_1}) reads
\begin{align}
\mathcal{K}^{\alpha}_j (\zi \omega_{\lambda})
	& = - \frac{2 \zi e J_1 J_2}{\beta V^2} \sum_{n} \iint \dd{\br} \dd{\br'}
\notag \\ & \hspace{-1em} \times
		\varphi_j (\zi \epsilon_n^{+}, \zi \epsilon_n; \br - \br')
		\left\{ \tilde{\bm{m}}_1 (\br) \times \tilde{\bm{m}}_2 (\br') \right\}^{\alpha}
\end{align}
with
\begin{align}
\varphi_j (\zi \epsilon_n^{+}, \zi \epsilon_n; \br)
	& = \sum_{\bq} \varphi_j (\zi \epsilon_n^{+}, \zi \epsilon_n; \bq) e^{\zi \bq \cdot \br}
\label{eq:varphi_1}
.\end{align}
By changing the variable $\bk + \bq \to \bq$ in Eq.~(\ref{eq:varphi_1}), we have
\begin{align}
\frac{1}{V} \varphi_j (\zi \epsilon_n^{+}, \zi \epsilon_n; \br)
	& = 2 g (\zi \epsilon_n^{+}; \br) \frac{\hbar}{\me \zi} \frac{\partial}{\partial r_j} Q (\zi \epsilon_n^{+}, \zi \epsilon_n; \br)
\notag \\ & \hspace{-2em}
		+ 2 g (\zi \epsilon_n; \br) \frac{\hbar}{\me \zi} \frac{\partial}{\partial r_j} Q (\zi \epsilon_n, \zi \epsilon_n^{+}; \br)
\notag \\ & \hspace{-2em}
		- 2 R (\zi \epsilon_n^{+}, \zi \epsilon_n; \br) \frac{\hbar}{\zi \me} \frac{\partial}{\partial r_j} R (\zi \epsilon_n^{+}, \zi \epsilon_n; \br)
,\end{align}
where we introduced the following notations,
\begin{align}
g (\zi \epsilon_n; \br)
	& \equiv \frac{1}{V} \sum_{\bk} g^{}_{\bk} (\zi \epsilon_n) e^{\zi \bk \cdot \br}
, \\
Q (\zi \epsilon_n^{+}, \zi \epsilon_n; \br)
	& \equiv \frac{1}{V} \sum_{\bk} \left( g_{\bk} (\zi \epsilon_n^{+}) \right)^2 g^{}_{\bk} (\zi \epsilon_n) e^{\zi \bk \cdot \br}
, \\
R (\zi \epsilon_n^{+}, \zi \epsilon_n; \br)
	& \equiv \frac{1}{V} \sum_{\bk} g_{\bk} (\zi \epsilon_n^{+}) g^{}_{\bk} (\zi \epsilon_n) e^{\zi \bk \cdot \br}
.\end{align}
By taking the analytic continuation $\zi \omega_{\lambda} \to \hbar \omega + \zi 0$ and assuming the absolute zero ($T = 0$), we can write down
\begin{align}
\frac{1}{\beta} \sum_n \varphi_j (\zi \epsilon_n^{+}, \zi \epsilon_n; \br)
	& = \varphi_j^{(0)} (\br)
		+ \zi \omega \varphi_j^{(1)} (\br)
		+ \cdots
\label{eq:after_ac}
,\end{align}
where only the $\omega$-linear term is of our interest.
The calculation detail of $\varphi_j^{(1)} (\br)$ is given in Appendix~\ref{apx:calculation_5layer}, the result of the calculation is shown as
\begin{align}
\frac{1}{V} \varphi_j^{(1)} (\br)
	& = \frac{\tau r_j}{\pi \hbar} \left[
		\left\{ g^{\R} (\br) \right\}^2 - \left\{ g^{\A} (\br) \right\}^2
	\right]
,\end{align}
where $\tau$ is the electron lifetime, $r_j$ is the $j$ component of the position $\br$; $r_x = x, r_y = y, r_z = z$, and $g^{\R/\A} (\br)$ is the retarded/advanced Green function in real space, which is given by
\begin{subequations}
	\begin{align}
g^{\R} (\br)
	& = - \frac{\me}{2 \pi \hbar^2} \frac{e^{\zi k_{F +} r}}{r}
, \\
g^{\A} (\br)
	& = - \frac{\me}{2 \pi \hbar^2} \frac{e^{- \zi k_{F -} r}}{r}
\end{align}
\label{eq:Green_function_r}%
\end{subequations}
with $k_{F \pm} = (\sqrt{2 \me}/\hbar) \sqrt{\mu \pm \zi \hbar / 2 \tau} = k_{F} \sqrt{1 \pm \zi / k_F l}$ with the mean free path $l = v_F \tau = \hbar k_F \tau / \me$.

Hence, the response coefficient $\chi_j^{\alpha} (\bq = 0)$ [Eq.~(\ref{eq:chi_q})] is obtained as
\begin{align}
\frac{1}{V} \chi_j^{\alpha} (\bq = 0)
	& = - 2 \zi e J_1 J_2 \int \frac{\dd{\br}}{V} \int \frac{\dd{\br'}}{V}
\notag \\ & \hspace{-1em} \times
		\varphi_j^{(1)} (\br - \br')
		\left\{ \tilde{\bm{m}}_1 (\br) \times \tilde{\bm{m}}_2 (\br') \right\}^{\alpha}
,\end{align}
which results in our desired expression of the damping-like Slonczewski torque,
\begin{align}
\bm{T}_i
	& = c J_i L_i \bm{M}_i \times (\bm{m}_1 \times \bm{m}_2)
\label{eq:T_i}
\end{align}
with $i = 1, 2$ and the coefficient $c$ given as
\begin{align}
c
	& = \frac{2 \zi e J_1 J_2 |\bm{E}| A}{\hbar} \int_{\Omega_1} \frac{\dd{\br}}{V} \int_{\Omega_2} \frac{\dd{\br'}}{V}
		\varphi_x^{(1)} (\br - \br')
\label{eq:c}
,\end{align}
where we assumed the specific configuration of Eq.~(\ref{eq:E}).
The integrals of $\br$ and $\br'$ can be done by using the assumption, $| \br - \br' | \simeq | x - x' |$.
The calculation detail is given in Appendix~\ref{apx:integral_5layer}, and the resultant expression is obtained as
\begin{align}
c J_i L_i
    & = \frac{3}{8 \pi} \frac{I_e}{e} \frac{A}{l_{sd, 1} l_{sd, 2}} \frac{L_i}{W} \frac{\Im [F]}{k_F l_{sd, i}}
\label{eq:result_c}
,\end{align}
where $I_e$ is the electric current,
\begin{align}
I_e
	& = A \sigma_e |\bm{E}|
	= G_e V_e
\end{align}
with the conductance $G_e = A \sigma_e / W$ and the voltage $V_e = W |\bm{E}|$, $k_F$ is the Fermi wavenumber, $l_{sd, i} = v_F \tau_{sd, i}$ is a typical length of the $sd$ exchange interaction with the typical time scale $\tau_{sd, i} = \hbar / 2 J_i$, and
\begin{align}
F
    & = \frac{- k_F}{2 \zi k_{F +}} e^{2 \zi k_{F +} \delta} \left( e^{2 \zi k_{F +} L_1} - 1 \right)  \left( e^{2 \zi k_{F +} L_2} - 1 \right)
\notag \\ & \hspace{1em}
        + k_F ( L_1 + L_2 + \delta) \mathrm{Ei} (2 \zi k_{F +} (L_1 + L_2 + \delta))
\notag \\ & \hspace{1em}
        - k_F ( L_1 + \delta) \mathrm{Ei} (2 \zi k_{F +} (L_1 + \delta))
\notag \\ & \hspace{1em}
        - k_F ( L_2 + \delta) \mathrm{Ei} (2 \zi k_{F +} (L_2 + \delta))
\notag \\ & \hspace{1em}
        + k_F \delta \mathrm{Ei} (2 \zi k_{F +} \delta)
\end{align}
with $L = L_1 + \delta$ and $\mathrm{Ei} (x)$ being the exponetial integral function.
Note that the electric conductivity $\sigma_e$ is given by $\sigma_e = e^2 \rho_e \tau / \me$ with $\rho_e = k_F^3 / 3 \pi^2$.

By taking the field theoretical approach, we successfully obtain the Slonczewski torque [Eq.~(\ref{eq:T_i})] with the coefficient given by Eq.~(\ref{eq:result_c}).
We discuss the result in Sec.~\ref{sec:discussion}.

\subsection{\label{sec:field-like}Field-like Slonczewski torque}
\begin{figure}[thpb]
\centering
\includegraphics[width=\linewidth]{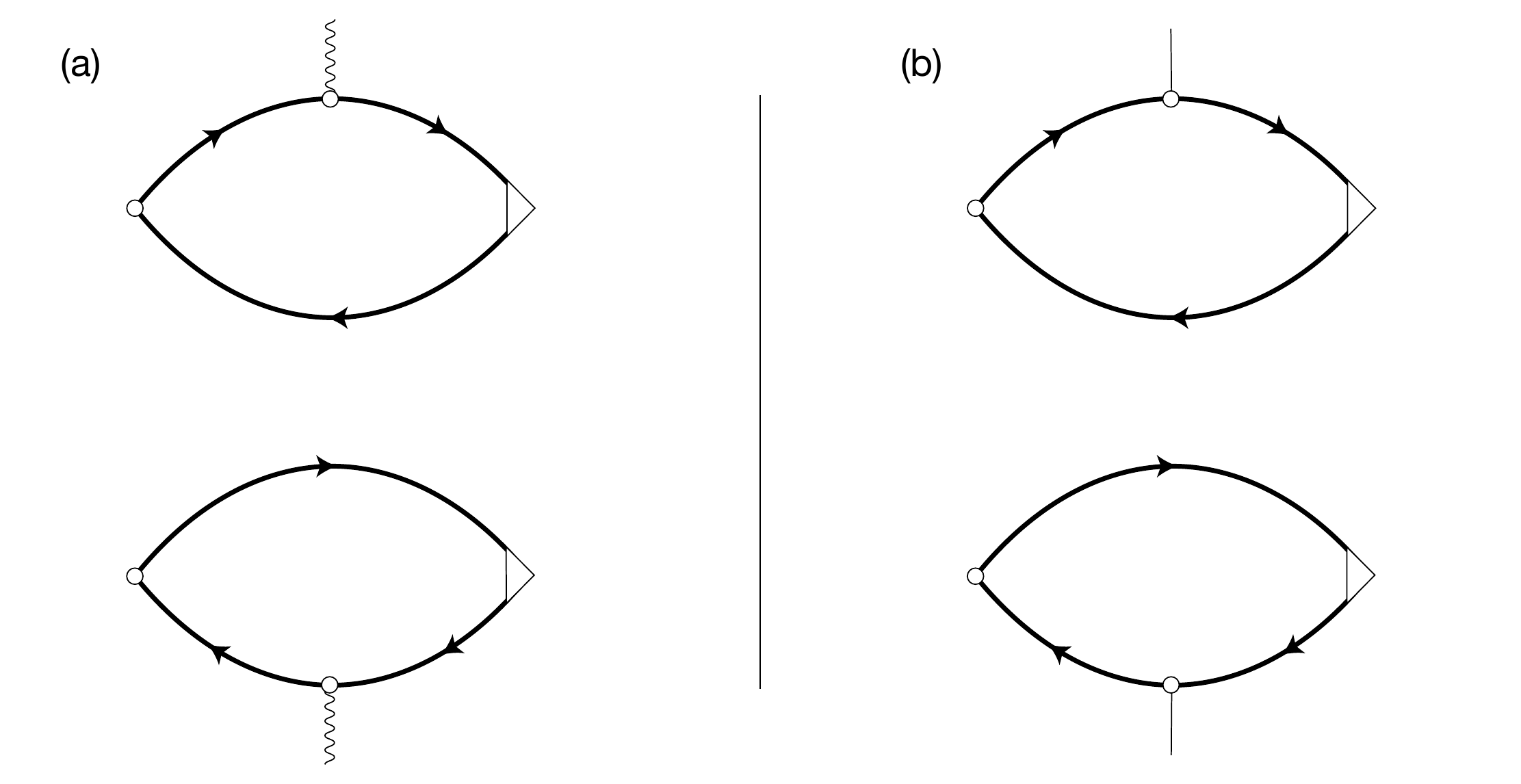}
\caption{\label{fig:diagrams_field-like}Feynman diagrams for the field-like Slonczewski torques on (a)~the magnetization $\bm{M}_1$ and (b)~the magnetization $\bm{M}_2$. The definitions of the lines and symbols are given in the caption of Fig.~\ref{fig:diagram_damping-like}. The key difference from Fig.~\ref{fig:diagram_damping-like} is that the spin vertex of the left sides changes the momentum, since we consider the response of the momentum-dependent spin.}
\end{figure}
Now, we evaluate the field-like Slonczewski torque, $\bm{T}'_1 \propto \bm{M}_1 \times \bm{m}_2$ and $\bm{T}'_2 \propto \bm{M}_2 \times \bm{m}_1$.
The field-like torque is also obtained through evaluating the electron spin [Eq.~(\ref{eq:response_s_in_k})] with Eq.~(\ref{eq:spin-torque}).
For the field-like Slonczewski torque, we expand the response coefficient up to the first order with respect to the $sd$-type exchange interactions, whose Feynman diagrams are given in Fig.~\ref{fig:diagrams_field-like}.
The response coefficient in this case reads
\begin{align}
\mathcal{K}_{j}^{\alpha} (\zi \omega_{\lambda}; \bq)
	& = - \frac{2 e}{\beta} \sum_{i} J_i \sum_n \vartheta_j (\zi \epsilon_n^{+}, \zi \epsilon_n; \bq) \tilde{m}_i^{\alpha} (-\bq)
\end{align}
with
\begin{align}
\vartheta_j (\zi \epsilon_n^{+}, \zi \epsilon_n; \bq)
	& = \frac{1}{V} \sum_{\bk} \left(
		g^{+}_{\bk+\bq}
		- g^{}_{\bk+\bq}
	\right) \frac{\hbar k_j}{\me} g^{+}_{\bk} g^{}_{\bk}
.\end{align}
In the real space representation, we get
\begin{align}
& \mathcal{K}_{j}^{\alpha} (\zi \omega_{\lambda}; \br)
\notag \\ 
	& = - \frac{2 e}{\beta V} \sum_{i} J_i \int \dd{\br'} \sum_n \vartheta_j (\zi \epsilon_n^{+}, \zi \epsilon_n; \br - \br') \tilde{m}_i^{\alpha} (\br')
,\end{align}
where 
\begin{align}
\vartheta_j (\zi \epsilon_n^{+}, \zi \epsilon_n; \br)
	& = \sum_{\bq} \vartheta_j (\zi \epsilon_n^{+}, \zi \epsilon_n; \bq) e^{\zi \bq \cdot \br}
,\end{align}
which is evaluated by taking the analytic continuation $\zi \omega_{\lambda} \to \hbar \omega + \zi 0$, and expand it as
\begin{align}
\frac{1}{\beta} \sum_n \vartheta_j (\zi \epsilon_n^{+}, \zi \epsilon_n; \br)
	& = \vartheta_j^{(0)} (\br)
		+ \zi \omega \vartheta_j^{(1)} (\br)
		+ \cdots
\label{eq:after_ac'}
,\end{align}
where the $\omega$-linear term is of our interest and given as
\begin{align}
\frac{1}{V} \vartheta_j^{(1)} (\br)
	& = \frac{- \hbar \tau}{2 \pi \me} 
		\frac{\partial}{\partial r_j} \left( g^{\R} (\br) - g^{\A} (\br) \right)^2
.\end{align}
From the above, we have
\begin{align}
\bm{T}'_1
	= c' \bm{M}_1 \times \bm{m}_2
, \qquad
\bm{T}'_2
	= - c' \bm{M}_2 \times \bm{m}_1
\label{eq:T'}
,\end{align}
where
\begin{align}
c'
	& = \frac{2 e J_1 J_2 |\bm{E}|}{\hbar V} \int_{\Omega_1} \dd{\br} \int_{\Omega_2} \dd{\br'} \vartheta_x^{(1)} (\br - \br')
.\end{align}
Here we have used $\vartheta_j^{(1)} (\br' - \br) = - \vartheta_j^{(1)} (\br - \br')$.

Substituting Eq.~(\ref{eq:Green_function_r}) into $\vartheta_j^{(1)} (\br)$, we finally obtain
\begin{align}
c'
	& = \frac{3}{16 \pi} \frac{I_e}{e} \frac{A}{l_{sd, 1} l_{sd, 2}} \{ h (L + L_2) - h (L) \}
,\end{align}
where we used the approximation $|\br - \br'| \simeq |x - x'|$, and
\begin{align}
h (y)
    & = \frac{1}{k_F} \int_0^{L_1} \dd{x} \frac{\left( e^{\zi k_{F +} (y - x)} - e^{- \zi k_{F -} (y - x)} \right)^2}{(y - x)^2}
\notag \\
    & = P (2 k_{F +}, y) - 2 P (k_{F +} - k_{F-}, y) + P (- 2 k_{F -}, y)
\end{align}
with
\begin{align}
P (k, y)
    & = \frac{1}{k_F} \int_0^{L_1} \dd{x} \frac{e^{\zi k (y - x)}}{(y - x)^2}
\notag \\
    & = \left[ \frac{e^{\zi k (y - x)}}{k_F (y - x)} - \frac{\zi k}{k_F} \mathrm{Ei} (k (y - x)) \right]_{x = 0}^{x = L_1}
\notag \\
    & = \frac{e^{\zi k (y - L_1)}}{k_F (y - L_1)} - \frac{\zi k}{k_F} \mathrm{Ei} (\zi k (y - L_1))
\notag \\ & \hspace{1em}   
    - \frac{e^{\zi k y}}{k_F y} + \frac{\zi k}{k_F} \mathrm{Ei} (\zi k y)
.\end{align}
We discuss the results in the following section.

\section{\label{sec:discussion}Results and Discussion}
Here, we discuss the obtained expressions of the damping-like and field-like Slonczewski torques.
Firstly, we successfully obtain the Slonczewski spin-transfer torque (damping-like torque) and the field-like torque corresponding to the $\beta$ torque in continuum systems, which indicates that our field-theoretical approach is valid for the spin-transfer spin torque in the magnetic junction system.
In the present calculation, we took the perturbation method of expanding the $sd$-type exchange interaction with the strength $J$ (to be exact, $J_1$ and $J_2$), and the damping-like and field-like torques are respectively proportional to $(J/\eF)^3$ and $(J/\eF)^2$, where $\eF$ is the Fermi energy.
Since the perturbation method should be valid only for the case of $J/\eF < 1$, the damping-like torque is always small than the field-like torque in our calculation.
For the case of strong $sd$-type exchange interaction, we need to take another method, such as the spin gauge field method.

Then, we see the coefficients of the obtained torques.
The first notable point is that the obtained coefficients seemingly do not contain the spin polarization of the electric current $P$, while Slonczewski showed that the damping-like torque depends on $P$ and vanishes when no spin polarization $P = 0$.
Since we expanded the $sd$-type exchange interaction, spin-dependent conductivity $\sigma_{s}$ ($s = \pm$) should be also expanded;
\begin{align}
\sigma_{s}
    & = \frac{e^2 \epsilon_{\mathrm{F} s} \nu_{s} \tau_{s}}{\me}
    \simeq \frac{2 s J_i}{\eF} \sigma_e
,\end{align}
where $\epsilon_{\mathrm{F} s} = \eF + s J_i$ is the Fermi energy, $\nu_s = \nu (\epsilon_{\mathrm{F} s}) \propto \sqrt{\epsilon_{\mathrm{F} s}}$ is the density of states, and the lifetime is given as $\hbar / \tau_s = 2 \pi n_i u^2 \nu_s$ with $n_i$ the impurity concentration and $u$ the impurity potential.
(Here, we omitted the index $i$ denoting the $i$-th magnetic layer with $i = 1, 2$ in $\epsilon_{\mathrm{F} s}$, $\nu_{s}$, $\tau_s$ and $\sigma_s$, for readability.) 
Hence, $\sigma_{\mathrm{s}} = \sigma_{+} - \sigma_{-} = P_i \sigma_e$ with $P_i = 4 J_i / \eF$ ($i = 1, 2$), which leads to the coefficient of the damping-like Slonczewski torque
\begin{align}
c J_1 L_1
    & = \frac{3}{24 \pi} \frac{P_{2} I_e}{e} \frac{A}{l_{sd, 1}^2} \frac{L_i}{W} \Im [F]
,\end{align}
although an ambiguity on which layer the spin polarization should be used remains.

The coefficient of the damping-like torque depends on the ferromagnetic layer, $\FI$ or $\FII$; $c J_1 L_1 \neq c J_2 L_2$, if $J_1 L_1 \neq J_2 L_2$, as seen in Eq.~(\ref{eq:T_i}).
It should be noted that the spin torque is given by Eq.~(\ref{eq:spin-torque}) and is defined as the integral of the electron spin density over the volume of the magnetic layer.
Since we have evaluated the uniform ($\bq = 0$) component of the electron spin for the damping-like torque, the integral over the volume of the magnetic layer is reduced to the volume of the magnetic layer.
On the other hand, the field-like torque has the same magnitude and opposite sign depending on the magnetic layer as seen in Eq.~(\ref{eq:T'}).
This critical feature shown in Eq.~(\ref{eq:T'}) does not change by considering the spin relaxation.
In continuum systems, the $\beta$ torques corresponding to the field-like torque arise from the spin-nonconserving process and nonadiabaticity.
Since our system is a spin-conserving model, the field-like torque might arise from the nonadiabaticity.
Note that the filed-like torque arises from the nonuniform ($\bq \neq 0$) component of the electron spin density.

\begin{figure*}[thb]
\includegraphics[width=\linewidth,clip]{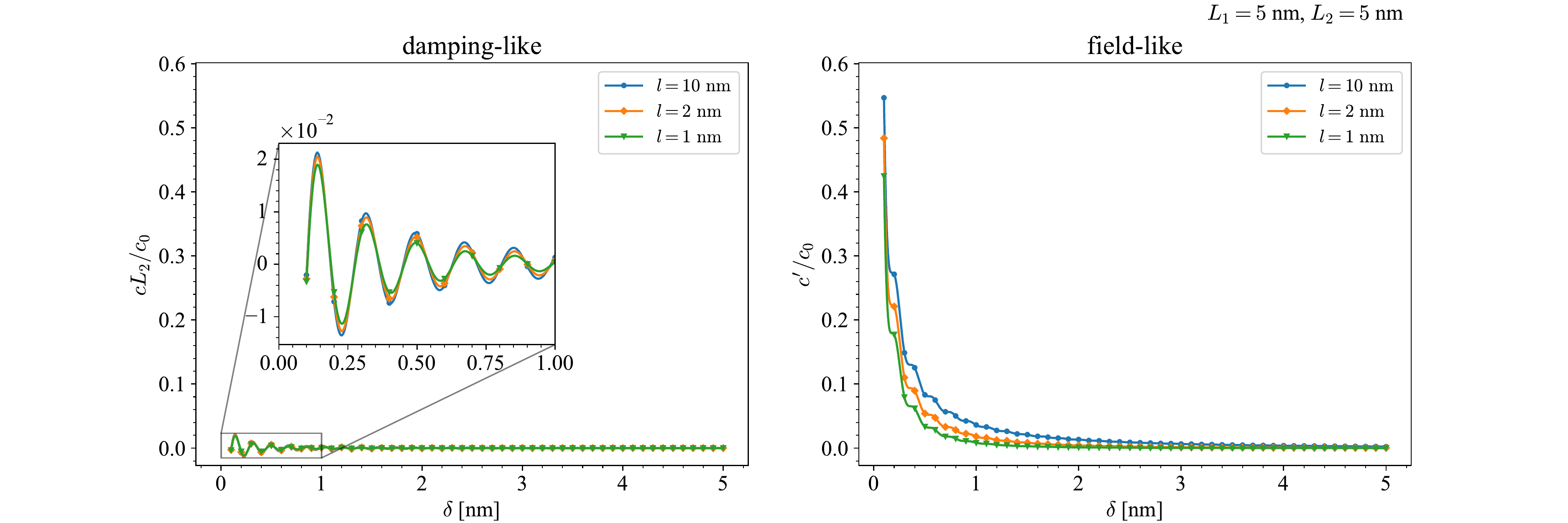}
\caption{\label{fig:4}~Dependence of the coefficients of the damping-like and field-like torques on the distance between the magnetic layers $\delta$ for various the mean free path $l$.
Both coefficients are measured by the unit $c_0 = (3 / 8 \pi) (I_e / e) (A / l_{sd,1} l_{sd,2})$.
We assume that the normal metal layer is made from aluminum; $k_F = 17.5~\mathrm{nm^{-1}}$, and set $k_F l_{sd,2} = J_2 / \eF = 0.5$, $W = L1 + L2 + \delta$, and $L_1 = L_2 = 5~\mathrm{nm}$.}
\end{figure*}
In Fig.~\ref{fig:4}, we depict the dependence of the damping-like and field-like torques on the distance between the magnetic layers $\delta$ with the fixed magnetic layer thicknesses $L_1 = L_2 = 5~\mathrm{nm}$ for various the mean free path $l$.
To compare the damping-like and field-like torques, we set $k_F l_{sd,2} = J_2 / \eF = 0.5$ and $W = L1 + L2 + \delta$, and measure the coefficients by the unit $c_0 = (3 / 8 \pi) (I_e / e) (A / l_{sd,1} l_{sd,2})$.
We assume that the normal metal layer is made from aluminum; $k_F = 17.5~\mathrm{nm^{-1}}$~\cite{takahashi2008}.
We see that both the torque coefficients decay as the distance between the magnetic layers increases, which is a natural result since the correlation of the magnetic layers is expected to decay.
The coefficient of the damping-like torque on the magnetization $\bm{M}_2$ is much smaller than that of the field-like torque in the entire region.
This relation in size is partial because the field-like torque is proportional to $(J/\eF)^2$ and the damping-like torque is in the order of $(J/\eF)^3$, as already mentioned.
Both the coefficients increase as the mean free path $l$ is larger, but the increments are slight.
For the damping-like torque, the coefficient, $c J_i L_i$, takes both the positive and negative values depending on the distance $\delta$.
This feature is shared with the Freidel oscillation and the Ruderman-Kittel-Kasuya-Yosida interaction.
On the other hand, the coefficient of the field-like torque, $c'$, takes only the positive values.

\begin{figure*}[thb]
\includegraphics[width=\linewidth,clip]{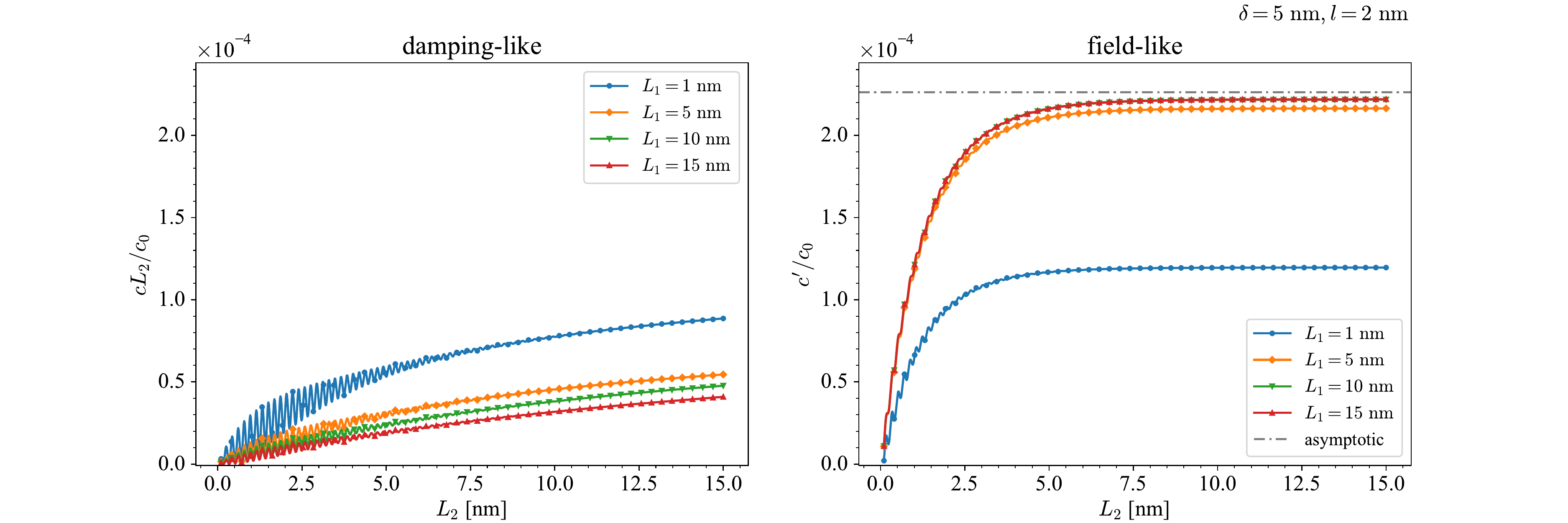}
\caption{\label{fig:5}~Dependence of the coefficients of the damping-like and field-like torques on $L_2$ for various $L_1$.
We set $\delta = 5~\mathrm{nm}$ and $l = 2~\mathrm{nm}$.
The unit $c_0$ and the parameters $k_F$, $k_F l_{sd,2}$, and $W$ are shown in the caption in Fig.~\ref{fig:4}.}
\end{figure*}
In Fig.~\ref{fig:5}, we show the dependence of the coefficients of the damping-like and field-like torques on $L_2$, the thickness of the magnetic layer $\FII$, with the distance $\delta = 5~\mathrm{nm}$ and the mean free path $l = 2~\mathrm{nm}$ for various thicknesses of the magnetic layer $\FI$.
We see that the coefficient of the damping-like torque increases slowly as $L_2$ increases, while that of the field-like torque increases more rapidly and saturates for $L_2 \gtrsim 5~\mathrm{nm}$.
Comparing the magnitudes of the damping-like and field-like torques, we confirm that the damping-like torque is smaller than the field-like torque, as discussed before.
Moreover, we find that the damping-like torque is smaller as $L_1$ is larger, while the field-like torque is larger, although the field-like torque saturates when $L_1 \gtrsim 10~\mathrm{nm}$.
Note that the sign of $c J_i L_i$ can change by changing $\delta$, while $c'$ is always positive, as seen in Fig.~\ref{fig:4}.

Here, we estimate the saturation value of $c'$ for $L_1, L_2 \gg l$.
The main contribution to $c'/c_0$ is from $- h (L_1 + \delta) / 2$, which is evaluated as
\begin{align}
c'
    & = \frac{c_0}{k_F} \int_{\delta / l}^{\infty} \dd{t} \frac{e^{-t} (1 - \cos 2 k_F l t)}{t^2}
\label{eq:c'_tmp}
,\end{align}
where we have used $\pm \zi k_{F \pm} \simeq - 1/2 l \pm \zi k_F$.
The term containing $\cos 2 k_F l t$ oscillates rapidly so that it cancels out mostly (but not completely).
Hence,
\begin{align}
c'
    & \simeq \frac{c_0}{k_F} \Gamma (-1, \delta/l)
\label{eq:c'_asymptotic}
,\end{align}
where $\Gamma (a, x) = \int_a^{\infty} \dd{t} t^{a-1} e^{-t}$ is the incomplete gamma function, which is expanded as
\begin{align}
\Gamma (-1, x)
    & = \left\{ \begin{array}{c c}
        1/x + \log x + \gamma - 1
    &   (x \ll 1)
    ,\\[1ex]
        e^{- x} / x^2
    &   (x \gg 1)
    .\end{array} \right.
\end{align}
Here, $\gamma$ is Euler's constant.
We also plot the asymptotic value~(\ref{eq:c'_asymptotic}) in Fig.~\ref{fig:5}, where the slightly difference between the asymptotic and exact values is seen, but the difference arises from the neglected term containing $\cos 2 k_F l t$ in Eq.~(\ref{eq:c'_tmp}).

\section{\label{sec:conclusion}Conclusion}
We have examined the field theoretical approach to the spin-transfer torques in the magnetic junction system composed of two ferromagnetic and three nonmagnetic metal layers.
We successfully obtain the damping-like and field-like Slonczewski torques by evaluating the nonequilibrium spin density due to the electric field.
The coefficient of the damping-like torque takes a different magnitude depending on the magnetizations in the magnetic layers, but that of the field-like torque has the same magnitude with the opposite sign in the magnetic layers.
We find that the coefficent of the damping-like torque has the spatial quantum oscillation like the Friedel oscillation, which has not been mentioned before.
We also find that the field-like torque has the saturation value for large thicknesses of the magnetic layers.

It is better to consider a more sophisticated model, such as a tight-binding model, which consists of the three domains; two ferromagnets and one nonmagnet, to describe the relation in size between the damping-like and field-like Slonczewski torques.

Since the field theoretical approach is valid for spin torques in magnetic junction systems, we are to examine the spin Hall torques based in the same way as we do in this work.
It will also be valuable to estimate the spin-torque ferromagnetic resonance based on the quantum field theory.

\begin{acknowledgments}
The author would like to thank M.~Hayashi, G.~Tatara, T.~Yamaguchi, and Y.~Araki.
This work is partially supported by JSPS KAKENHI Grant Number JP22K13997.
\end{acknowledgments}
\appendix

\section{\label{apx:calculation_5layer}Calculation of damping-like Slonczewski torque}
Here, we give the calculation detail of the damping-like Slonczewski torque.
The $\omega$-linear term in Eq.~(\ref{eq:after_ac}) is obtained as
\begin{align}
\frac{\pi}{\hbar} \varphi_i^{(1)} (\br)
	& = g^{\R} (\br) \frac{\hbar}{\zi \me} \frac{\partial}{\partial r_i} Q^{\R \A} (\br)
\notag \\ & 
	+ g^{\A} (\br) \frac{\hbar}{\zi \me} \frac{\partial}{\partial r_i} Q^{\A \R} (\br)
	\notag \\ & 
	- R^{\R \A} (\br) \frac{\hbar}{\zi \me} \frac{\partial}{\partial r_i} R^{\R \A} (\br)
,\end{align}
where we have defined
\begin{align}
g^{\X} (\br)
	& = \frac{1}{V} \sum_{\bk} g^{\X}_{\bk} e^{\zi \bk \cdot \br}
 \quad (\X \in \{ \R, \A \})
, \\
Q^{\X \Y} (\br)
	& = \frac{1}{V} \sum_{\bk} \left( g_{\bk}^{\X} \right)^2 g^{\Y}_{\bk} e^{\zi \bk \cdot \br}
 \quad (\X, \Y \in \{ \R, \A \})
, \\
R^{\R \A} (\br)
	& = \frac{1}{V} \sum_{\bk} g^{\R}_{\bk} g^{\A}_{\bk} e^{\zi \bk \cdot \br}
,\end{align}
and $g^{\R/\A}_{\bk}$ is the retarded/advanced Green function
\begin{align}
g^{\R/\A}_{\bk}
	& = \frac{1}{\mu - \frac{\hbar^2 k^2}{2 \me} \pm \frac{\zi \hbar}{2 \tau}}
.\end{align}
Note that we introduced the lifetime of the electron $\tau$.

We use the following relation,
\begin{align}
g^{\R}_{\bk} g^{\A}_{\bk}
	& = \frac{1}{\left( \mu + \frac{\zi \hbar}{2 \tau} - \frac{\hbar^2 k^2}{2 \me} \right) \left( \mu - \frac{\zi \hbar}{2 \tau} - \frac{\hbar^2 k^2}{2 \me} \right)}
\notag \\
	& = - \frac{\tau}{\zi \hbar} \left(
		\frac{1}{\mu + \frac{\zi \hbar}{2 \tau} - \frac{\hbar^2 k^2}{2 \me}}
		- \frac{1}{\mu - \frac{\zi \hbar}{2 \tau} - \frac{\hbar^2 k^2}{2 \me}}
	\right)
\notag \\
	& = - \frac{\tau}{\zi \hbar} \left( g^{\R}_{\bk} - g^{\A}_{\bk} \right)
\end{align}
and obtain
\begin{align}
Q^{\R \A} (\br)
	& = - \frac{\tau}{\zi \hbar} \frac{1}{V} \sum_{\bk} \left\{
		\left( g_{\bk}^{\R} \right)^2
		- g^{\R}_{\bk} g^{\A}_{\bk}
	\right\} e^{\zi \bk \cdot \br}
\notag \\
	& = - \frac{\tau}{\zi \hbar} \frac{1}{V} \sum_{\bk} \left( g_{\bk}^{\R} \right)^2 e^{\zi \bk \cdot \br}
		+ \frac{\tau}{\zi \hbar} R^{\R \A} (\br)
.\end{align}
Similarly, 
\begin{align}
Q^{\A \R} (\br)
	& = - \frac{\tau}{\zi \hbar} \frac{1}{V} \sum_{\bk} \left\{
		g^{\R}_{\bk} g^{\A}_{\bk}
		- \left( g_{\bk}^{\A} \right)^2
	\right\} e^{\zi \bk \cdot \br}
\notag \\
	& = \frac{\tau}{\zi \hbar} \frac{1}{V} \sum_{\bk} \left( g_{\bk}^{\A} \right)^2 e^{\zi \bk \cdot \br}
		- \frac{\tau}{\zi \hbar} R^{\R \A} (\br)
,\end{align}
which results in
\begin{align}
& g^{\R} (\br) \frac{\hbar}{\zi \me} \frac{\partial}{\partial r_i} Q^{\R \A} (\br)
	+ g^{\A} (\br) \frac{\hbar}{\zi \me} \frac{\partial}{\partial r_i} Q^{\A \R} (\br)
\notag \\
	& = g^{\R} (\br) \frac{\hbar}{\zi \me} \frac{\partial}{\partial r_i} \left(
		- \frac{\tau}{\zi \hbar} \frac{1}{V} \sum_{\bk} \left( g_{\bk}^{\R} \right)^2 e^{\zi \bk \cdot \br}
	\right)
\notag \\ & \hspace{1em}
	+ g^{\A} (\br) \frac{\hbar}{\zi \me} \frac{\partial}{\partial r_i} \left(
		\frac{\tau}{\zi \hbar} \frac{1}{V} \sum_{\bk} \left( g_{\bk}^{\A} \right)^2 e^{\zi \bk \cdot \br}
	\right)
\notag \\ & \hspace{1em}
	+ \frac{\tau}{\zi \hbar} \left\{ g^{\R} (\br) - g^{\A} (\br) \right\}
		\frac{\hbar}{\zi \me} \frac{\partial}{\partial r_i} R^{\R \A} (\br)
.\end{align}
Note that the last term can be rewritten as
\begin{align}
\frac{\tau}{\zi \hbar}
& \left\{ g^{\R} (\br) - g^{\A} (\br) \right\} 
	\frac{\hbar}{\zi \me} \frac{\partial}{\partial r_i} R^{\R \A} (\br)
\notag \\
	& = - R^{\R \A} (\br) \frac{\hbar}{\zi \me} \frac{\partial}{\partial r_i} R^{\R \A} (\br)
.\end{align}
Hence,
\begin{align}
\frac{\pi}{\hbar} \varphi_i^{(1)} (\br)
	& = g^{\R} (\br) \frac{\hbar}{\zi \me} \frac{\partial}{\partial r_i} \left(
		- \frac{\tau}{\zi \hbar} \frac{1}{V} \sum_{\bk} \left( g_{\bk}^{\R} \right)^2 e^{\zi \bk \cdot \br}
	\right)
\notag \\ & \hspace{1em}
	+ g^{\A} (\br) \frac{\hbar}{\zi \me} \frac{\partial}{\partial r_i} \left(
		\frac{\tau}{\zi \hbar} \frac{1}{V} \sum_{\bk} \left( g_{\bk}^{\A} \right)^2 e^{\zi \bk \cdot \br}
	\right)
.\end{align}
Moreover,
\begin{align}
\frac{\hbar}{\zi \me} \frac{\partial}{\partial r_i} \left(
	\frac{\tau}{\zi \hbar} \frac{1}{V} \sum_{\bk} \left( g_{\bk}^{\X} \right)^2 e^{\zi \bk \cdot \br}
\right)
	& = \frac{\tau}{\zi \hbar} \frac{1}{V} \sum_{\bk} \frac{\hbar k_i}{\me} \left( g_{\bk}^{\X} \right)^2 e^{\zi \bk \cdot \br}
\notag \\
	& = - \frac{\tau}{\zi \hbar^2} \frac{1}{V} \sum_{\bk} g_{\bk}^{\X} \frac{\partial}{\partial k_i} e^{\zi \bk \cdot \br}
\notag \\
	& = - \frac{\tau r_i}{\hbar^2} g^{\X} (\br)
,\end{align}
where we have used $\partial_{k_i} g^{\X}_{\bk} = (\hbar^2 k_i / \me) (g^{\X}_{\bk})^2$.

We finally obtain
\begin{align}
\varphi_i^{(1)} (\br)
	& = \frac{\hbar}{\pi} \frac{\tau r_i}{\hbar^2} \left[
		\left\{ g^{\R} (\br) \right\}^2 - \left\{ g^{\A} (\br) \right\}^2
	\right]
.\end{align}

\section{\label{apx:integral_5layer}Integrals in damping-like Slonczewski torque}
Here, we show the calculation of the integrals in Eq.~(\ref{eq:c});
\begin{align}
c
  & = \frac{\zi \me^2 e J_1 J_2 |\bm{E}| A \tau}{2 \pi^3 \hbar^6 A W} \int_{\Omega_1} \dd{\br}\int_{\Omega_2} \dd{\br'}
\notag \\ & \hspace{3em} \times
    \frac{(x- x')}{|\br - \br'|^2} \left[
		e^{2 \zi k_{F +} |\br - \br'|} - e^{- 2 \zi k_{F -} |\br - \br'|}
	\right]
.\end{align}
We presume that the part $|\br - \br'| \simeq |x - x'|$ plays the important role, so that we have
\begin{align}
c J_i L_i
    & = \frac{3}{8 \pi} \frac{I_e}{e} \frac{A}{l_{sd,1} l_{sd,2}} \frac{L_i}{W} \frac{\Im [F]}{k_F l_{sd, i}}
\end{align}
with
\begin{align}
F
    & = k_F \int_{0}^{L_1} \dd{x} \int_{L}^{L+L_2} \dd{x'}
        \frac{e^{2 \zi k_{F +} (x' - x)}}{x' - x}
.\end{align}
We can evaluate $F$ as
\begin{align}
F
    & = k_F \int_{0}^{L_1} \dd{x} \int_{L - x}^{L+L_2 - x} \dd{t} \frac{e^{2 \zi k_{F +} t}}{t}
\notag \\
    & = k_F \int_{0}^{L_1} \dd{x} \left[
            \mathrm{Ei} (2 \zi k_{F +} (L + L_2 - x))
            - \mathrm{Ei} (2 \zi k_{F +} (L - x))
        \right]
\notag \\
    & = k_F \frac{- 1}{2 \zi k_{F +}} \left( e^{2 \zi k_{F +} L} - e^{2 \zi k_{F +} (L - L_1)} \right)  \left( e^{2 \zi k_{F +} L_2} - 1 \right)
\notag \\ & \hspace{1em}
        + k_F ( L + L_2) \mathrm{Ei} (2 \zi k_{F +} (L + L_2))
\notag \\ & \hspace{1em}
        - k_F L \mathrm{Ei} (2 \zi k_{F +} L)
\notag \\ & \hspace{1em}
        - k_F ( L + L_2 - L_1) \mathrm{Ei} (2 \zi k_{F +} (L + L_2 - L_1))
\notag \\ & \hspace{1em}
        + k_F (L - L_1) \mathrm{Ei} (2 \zi k_{F +} (L - L_1)))
,\end{align}
weher $\mathrm{Ei} (x)$ is the exponetial integral function.
Here, introducing $L = L_1 + \delta$, we obtain
\begin{align}
F
    & = \frac{- k_F}{2 \zi k_{F +}} e^{2 \zi k_{F +} \delta} \left( e^{2 \zi k_{F +} L_1} - 1 \right)  \left( e^{2 \zi k_{F +} L_2} - 1 \right)
\notag \\ & \hspace{1em}
        + k_F ( L_1 + L_2 + \delta) \mathrm{Ei} (2 \zi k_{F +} (L_1 + L_2 + \delta))
\notag \\ & \hspace{1em}
        - k_F ( L_1 + \delta) \mathrm{Ei} (2 \zi k_{F +} (L_1 + \delta))
\notag \\ & \hspace{1em}
        - k_F ( L_2 + \delta) \mathrm{Ei} (2 \zi k_{F +} (L_2 + \delta))
\notag \\ & \hspace{1em}
        + k_F \delta \mathrm{Ei} (2 \zi k_{F +} \delta)
\end{align}

\bibliography{stt}
\end{document}